Revealing the nanoparticle composition of Edvard Munch's *The Scream*, and implications for paint alteration in iconic early 20th century artworks.

Barnaby D.A. Levin[1], Adam C. Finnefrock[2], Alyssa M. Hull[2], Malcolm G. Thomas[3], Kayla X. Nguyen[1], Megan E. Holtz[1], Unn Plahter[4], Inger Grimstad[5], Jennifer L. Mass[2,6] & David A. Muller[1,7]

1. School of Applied and Engineering Physics, Cornell University, Ithaca, NY, USA.
2. Scientific Analysis of Fine Art LLC, New York, NY, USA
3. Cornell Center for Materials Research, Cornell University, Ithaca, NY, USA
4. University of Oslo, Oslo, Norway.
5. Munch, Museum, Oslo, Norway.
6. Bard Graduate Center, New York, NY, USA
7. Kavli Institute for Nanoscale Science, Cornell University, Ithaca, NY, USA.

**A major motivation for the scientific study of artworks is to understand their states of preservation and ongoing degradation mechanisms. This enables preservation strategies to be developed for irreplaceable works. Intensely-hued cadmium sulphide (CdS) yellow pigments are of particular interest because these are key to the palettes of many important late 19th and early 20th century masters, including Vincent Van Gogh, Pablo Picasso, Henri Matisse, and Edvard Munch. As these paintings age, their cadmium yellow paints are undergoing severe fading, flaking, and discolouration.[1-5] These effects are associated with photodegradation, the light-facilitated reactions of CdS with oxygen, moisture, and even the paint binding medium.[5-12] The use of common optical and X-ray methods to characterize the physical state of the pigment is challenging due to the mixing of the various components of the paint at length scales smaller than their resolution.**

**Here we present an atomic-scale structural and chemical analysis of the CdS pigment in Edvard Munch's *The Scream* (c.1910, Munch Museet) enabled by new electron microscope detector technologies. We show that the CdS pigment consists of clusters of defective nanoparticles ~5-10 nm in diameter. It is known from the modern use of such particles in photocatalysis that they are inherently vulnerable to photodegradation[13-16]. Chlorine doping and a polytype crystal structure further enhance the sensitivity of the CdS pigment to photodegradation. In addition to *The Scream*, we have also observed this inherently unstable pigment structure in Henri Matisse's *Flower Piece* (1906, Barnes Foundation). The fundamental understanding of the pigments' nanoscale structures and impurities described here can now be used to predict which paintings are most at risk of photo-oxidation and guide the most effective preservation strategies for iconic masterpieces.**

Figure 1a shows the location on *The Scream* (c. 1910, Munch Museum) from which the microsample of paint analysed in this study was originally extracted in 1974 and kept in dark storage. Previous analyses of cadmium yellow paint degradation in *The Scream* and contemporary paintings by optical, SEM, X-ray emission, and X-ray absorption techniques have shown that the paint from this period has a complex microscale structure, consisting of a number of different species of cadmium compounds. Recent work has highlighted the roles of high relative humidity and zinc substitution in the crystal lattice in this degradation[3-5]. Alongside the CdS pigment, cadmium carbonate ($CdCO_3$), a white compound and potential photodegradation product or residual synthesis reagent, has been found to be a major component[6-11]. However, the different species are intermixed at the microscale, making detailed analysis of individual species and their relationships to each other challenging. The unique advantage of TEM is that it allows us to characterize paint particles at the nanoscale, giving us information about particle size, morphology, structure, composition, and the distribution of different species of cadmium compounds. We used a focused ion beam (FIB) to extract two thin cross-sections[17], referred to as "Scream 1" and "Scream 2" from two ~40 μm-wide flakes of paint in the microsample for nanoscale TEM characterization (Figure 1b-d, Extended Data Figure 1).

Spectroscopic analysis of the paint cross sections shows that only a small proportion of paint consists of CdS pigment particles, and that CdS is concentrated in small regions of ~ 50-500 nm in diameter (Figure 2). We have found that a large proportion of paint particles consist of $CdCO_3$ (Figure 2a-b) by using electron energy loss spectroscopy (EELS) mapping to inspect the fine structure of the carbon K-edge in the paint cross sections, which is consistent with bulk measurements[6-11]. We have also detected chlorine in some of the particles in the paint cross sections using X-ray energy dispersive spectroscopy (XEDS) mapping. Some particles contain cadmium and chlorine, but little to no oxygen or sulphur (the blue regions of Figure 2d), suggesting that they are cadmium chloride particles. Additionally, X-ray spectra from the CdS particles show a small chlorine peak, suggesting that the CdS particles are n-type doped with chlorine atoms. Example spectra are shown in Extended Data Figures 2 and 3, along with additional elemental maps. This is particularly significant because n-type doping of CdS is known to increase its sensitivity to photodegradation[14,15], as is the presence of trap states[12], which may be caused by doping. Chlorine doping may therefore be an important factor governing the extent of photodegradation of cadmium yellow paint in *The Scream*.

The structure of regions of CdS pigment was examined using high magnification imaging. Figure 3a shows a high angle annular dark field (HAADF) scanning TEM (STEM) image of a region of CdS particles in sample Scream 1. The bright regions of this image correspond to CdS, and have a granular appearance, which suggests that the CdS pigment consists of clusters of small nanoparticles, which we estimate to be of ~5-10 nm in size. This is further supported by energy-filtered TEM (EFTEM) imaging, which allows us to image individual nanoparticles with lattice resolution (Figure 3b). Convergent beam electron diffraction (CBED) mapping suggests that all of the CdS in both of our TEM samples share the same nanoparticle structure. Using information from the diffraction patterns, we can construct a map showing the locations of all of the larger crystalline grains in our samples (this process is described further in Methods), and overlay this with the spectroscopic map corresponding to cadmium sulphide (Figure 3c). We observe no clear overlap between these large crystal grains and regions of cadmium sulphide anywhere in either sample. Inspection of individual CBED patterns from clusters of CdS shows a polycrystalline ring-like pattern, which is consistent with the nanoparticle structure that we observe in our high magnification images given the sub-nm probe size that we used for CBED mapping. Indeed, given that at least 10 nanoparticles are likely required to produce a ring pattern, and given the ~ 150 nm thickness of our cross section, we can place an upper limit of ~15 nm on the CdS particle size. These observations that the CdS pigment consists of small nanoparticles are a key result because nanoparticles are typically much more reactive than bulk materials due to their relatively large surface area to volume ratio. The nanoparticle structure of the CdS pigment is therefore likely to be a significant factor that increases the sensitivity of the cadmium yellow paint to photodegradation. We note that quantum confinement effects are unlikely to play a role, as this effect is weak for particles > 5 nm[18-19].

The crystal structure of the CdS pigment particles is another factor that may contribute to the photosensitivity of the cadmium yellow paint. CdS is known to have two stable crystal phases in bulk. Greenockite, or α-CdS, has a hexagonal wurtzite structure, and hawleyite, or β-CdS, has a cubic zinc-blende structure.[20] Stacking faults between cubic and hexagonal domains within the CdS nanoparticles can give rise to a polytype crystal structure.[20-22] The inset in Figure 3b shows one example of a varying atomic stacking sequence in a nanoparticle, which suggests that the CdS particles in *The Scream* may have a polytype structure. This is further supported by selected area electron diffraction (SAED) patterns from clusters of CdS nanoparticles (Figure 4), and Fourier transforms of lattice resolution EFTEM images acquired over large fields of view (Extended Data Figure 7). The rotational average of the SAED patterns (R-SAED) gives equivalent information to a powder X-ray diffraction pattern, allowing us to compare the crystal structure of the CdS nanoparticles in *The Scream* to that those reported in the literature. The R-SAED patterns of CdS pigment particles in *The Scream* show broad peaks that can be indexed to either the greenockite or hawleyite structures (Figure 4c). Significantly, the hawleyite (200) peak and greenockite (102) peaks are either very weak, or absent from these patterns, which is typical when particles have a polytype crystal structure[21]. The polytypism of the CdS nanoparticles may be a third factor governing the extent of photodegradation of cadmium yellow paint in *The Scream*, because polytype nanoparticles can exhibit complex morphologies, with a higher surface area to volume ratio than monotype nanoparticles.[23] This is also an important observation because it implies that attempts to estimate CdS particle size using X-ray diffraction may give an incorrect result if only the greenockite or hawleyite structures are assumed.

In addition to providing new insights into the factors affecting the photosensitivity of cadmium yellow paint in *The Scream*, our observations may also provide insight into aspects both of the paint synthesis method, and of the CdS photodegradation mechanism. Three potential synthesis methods have been discussed;[7] a dry synthesis method, an indirect wet synthesis method, and a direct wet synthesis method. Our observation that the CdS pigment in *The Scream* consists of polytype nanoparticles is consistent with an indirect wet synthesis method because this method is known to produce CdS nanoparticles of a similar size and structure.[24-25] The presence of chlorine in the paint is also consistent with a wet synthesis method, because cadmium chloride can be used as a starting reagent in wet CdS synthesis[4]. $CdCO_3$ may be the end product of CdS photodegradation.[9] Indeed, the conversion of yellow CdS into white $CdCO_3$ is consistent with the fact that the paint has been observed to whiten with age[9-11]. With electron diffraction and high resolution EFTEM imaging, we have observed two $CdCO_3$ particle sizes (Extended Data Figure 9). The larger ~1 μm particles may have originally been present in the paint; smaller ~10 nm particles, similar in size to the CdS nanoparticles are likely to be CdS aging products (see Supplementary discussion).

Having determined that the CdS pigment in *The Scream* was composed of chlorine-doped polytype nanoparticles, we sought to determine whether similar structures may be found in other works from the period. To this end, we analyzed CdS pigment particles in microtomed sections of paint from *Flower Piece* by Henri Matisse (see Extended Data Figures 10-12 and Supplementary Discussion). We discovered that the structure of CdS particles in *Flower Piece* was indeed very similar to those in *The Scream*, suggesting that CdS particle size, polytype crystal structure, and chlorine content are also key factors influencing the rate of photodegradation of cadmium yellow paint in *Flower Piece*. Previous research by some of these authors has shown cadmium yellow degradation to also be present in Henri Matisse's *Le Bonheur de vivre* (1905-6, Barnes Foundation)[9-11] as well as works by a number of Edvard Munch and Henri Matisse's contemporaries[6,7], including Vincent Van Gogh[8].

Research on cadmium yellow pigments in the works of the early modernists has previously focused on phase identification and implications for the mechanisms of degradation. This has been essential for recommending appropriate display conditions for these works. The study presented here, however, leads to several new methods for predicting which paintings are most likely to undergo degradation. These include works with cadmium yellow pigments that are nanocrystalline, that have chlorides present in the CdS crystal lattice, and that have crystallized into the polytype form of the compound. With this information, conservation scientists and museum professionals can devote the most rigorous preventive conservation protocols to the works by early modernist artists, including Edvard Munch and Henri Matisse, that are most likely to deteriorate in the future.

**Methods**

**FIB.** In our study of paint from *The Scream*, we used a focused ion beam (FIB) for sample preparation to preserve the microstructure of the sample paint that we extracted, so that we could observe the distribution of different chemical components within the paint. Cross sections of paint were prepared using an FEI Strata 400S FIB system equipped with an Omniprobe 200 needle and lift out system. Particles were selected for FIB sectioning based on the detection of both cadmium and sulphur in XEDS spectra of the paint flakes acquired in the FIB (not shown), and the presence of a relatively flat surface for the deposition of a protective platinum layer. A ~2 μm thick layer of platinum was deposited in a ~20 μm x 2 μm rectangular area, and paint surrounding this area was milled away with the ion beam, allowing an initial ~2 μm thick paint cross section to be extracted from the paint flake. This initial cross section was then attached to a copper half-grid and thinned further with the ion beam. The FIB process is illustrated in Extended Data Figure 1 below. Two cross sections of paint were prepared in this manner, which we refer to as "Scream 1" (Extended Data Figure 1 g-h) and "Scream 2" (Extended Data Figure 1i-j). The final milling step was conducted at an ion beam energy of 5 keV to minimize damage from the gallium ion beam. XEDS spectra were taken from the cross section prior to final thinning to confirm the presence of both cadmium and sulphur. Typically, FIB is used to thin TEM samples to <40 nm to ensure a high degree of electron transparency for high resolution imaging. However, paint samples were observed to begin to bend and curl as they were thinned down to ~100-150 nm (Extended Data Figure 1h). We attribute this to the low elastic modulus of the binder that holds the paint medium together. Additionally, pores or voids in the 3D structure of the paint flakes will become holes in a thin cross section, further reducing the structural integrity of FIB thinned paint samples. Sample Scream 1 was thinned until a bend in the cross section was observed, and sample Scream 2 was thinned until a ~500 nm hole in the structure was observed

**XEDS.** XEDS maps of the sample were acquired with an Oxford Instruments XMAX detector on an FEI Tecnai F-20 S/TEM operated at 200 kV, in STEM mode. Since a relatively large (~2 nA) probe current is required to generate a strong XEDS signal on our instrument, XEDS maps were acquired after all of the other data had been collected in order to avoid any potential artefacts due to radiation damage. In our spectra, the sulphur K$\alpha$ edge partly overlapped with the platinum M$\beta$ edge from the platinum capping layer deposited during the FIB procedure described above. Spatial maps showing only the sulphur K$\alpha$ edge signal were generated by using a least-squares fitting procedure in MATLAB, with a pure sulphur K$\alpha$ edge, and a pure platinum M$\beta$ edge as reference spectra.

**EELS.** EELS maps were acquired on an FEI Titan Themis S/TEM, operated at 300 kV in STEM mode. The microscope was equipped with a Gatan Enfinium Spectrometer, and an energy dispersion of 0.25 eV per channel was used. In sample Scream 1, our thinnest sample, we were able to collect EELS datasets from the entire thinned region. A high voltage, 300 kV electron beam was used in order to reduce plural scattering effects, and minimize the background due to the thickness of the sample. As a wide area of the sample was mapped, simultaneous low loss and core loss EELS datasets were acquired, and the position of the zero-loss peak in the low loss spectra was used to align the core loss data. Alignment of spectra was performed using custom MATLAB code. The core loss spectra were collected over an energy range of 110-620 eV. Sub-pixel scanning was used to minimize the dwell time of the electron beam at any one probe position, to reduce the risk of radiation damage to the sample. When analyzing the carbon K-edge, separate binder and $CdCO_3$ maps are generated using a least squares method in MATLAB to fit the different carbon K-edges to the data. The fine structure of the carbon K-edge for the carbonate edge was readily identified and distinguished from that of the binding medium by the presence of two sharp peaks[26].

**EFTEM.** Zero loss filtered EFTEM images were acquired using an FEI Titan Themis S/TEM operated at 300 kV in TEM mode, equipped with a Gatan GIF Tridiem energy filter, using a 20 eV wide energy slit. Images were recorded on a 3710 x 3838 pixel Gatan K2 camera, which allowed us to acquire lattice resolution images with hundreds of CdS nanoparticles in a single field of view.

**SAED.** Selected Area Electron Diffraction (SAED) patterns were acquired from the sample using an FEI Tecnai F-20 S/TEM operated at 200 kV, in TEM mode. Diffraction patterns were recorded on a Gatan Orius CCD camera.

**EMPAD.** We used a recently developed Electron Microscope Pixel Array Detector (EMPAD)[27] to acquire a convergent beam electron diffraction (CBED) maps of our samples. The EMPAD datasets for sample Scream 1 were acquired on an FEI Titan Themis S/TEM, operated at 300 kV in STEM mode. EMPAD datasets for sample Scream 2 were acquired on an FEI Tecnai F-20 S/TEM operated at 200 kV in STEM mode. EMPAD datasets are 4-dimensional, because each pixel of a 2-dimensional map contains a 2-dimensional CBED pattern. A ~1 nm probe was used, which allowed us to record CBED patterns in which diffraction disks did not overlap. Under these conditions, we estimated that a CBED pattern will appear to contain polycrystalline rings if there are more than 10 crystallites in the path of the beam. Given our probe size and sample thickness (> 100 nm), polycrystalline rings in the CBED pattern were therefore taken to indicate a particle size of ~ 10 nm or less.

We were able to map the positions of all larger (>10 nm) particles in our samples by generating centre of mass images of the sample using the method of Caswell et al[28]. This method locates single crystals by mapping positions which have an asymmetric diffraction pattern. This is performed over a small range of tilts to ensure that on-axis crystal grains are included in the map. Amorphous material will not appear in centre of mass maps because the CBED patterns of amorphous materials are diffuse and featureless, and therefore essentially symmetric. Similarly, nanocrystalline material (crystal size <10 nm) will not appear in centre of mass maps because the CBED patterns from nanocrystallites have symmetric polycrystalline rings. Experimentally, we observed that the centre of mass method does not always capture very large (~1 µm) sized crystal grains, when only a single tilt axis is available. However, these large crystal grains can be directly mapped by generating dark-field images from the dataset using virtual apertures placed over their diffraction spots in the CBED patterns, and subtracting the background by using virtual apertures placed to the side of their diffraction spots (examples are shown in Extended Data Figure 9).

**Ultra-microtome**. We chose to test an alternative, ultra-microtome sample preparation method for paint from Heri Matisse's *The Flower Piece*. A flake of paint was embedded in an epoxy and sectioned using a Leica Ultracut UCT Ultra-microtome to a thickness of ~50 nm. An advantage of the ultra-microtome method is that samples containing relatively large quantities of material could be prepared for TEM analysis with a high through-put compared to the FIB method. However, we found that a disadvantage of the technique is that paint particles in the sample can move as the microtomed section is cut, altering the microscale distribution of particles. Furthermore, particles can become dislodged and fall out of very thin microtomed sections, meaning that some components of the paint may be missing in the electron microscope analysis. Despite the disadvantages of the technique, CdS particles remained present, and we were able to characterize the nanostructure of the CdS pigment, (Extended Data Figures 10-12). We concluded that for future studies, the FIB method may be more appropriate because the position and distribution of particles in the sample at the microscale is preserved more effectively using FIB. In future studies, we will aim to apply the FIB technique to cross sections of paint from *The Flower Piece* to allow us to investigate the structure and composition of the paint in more detail.

Acknowledgements

The authors thank Prof. Hector Abruña, Cornell University Department of Chemistry for useful discussions regarding CdS degradation. This work made use of the Cornell Center for Materials Research Facilities supported by the National Science Foundation under Award Number DMR-1719875. We thank John Grazul and Mariena Silvestry Ramos for assistance in the TEM facilities. We acknowledge the Munch Museum providing a paint sample and for use of their image of *The Scream,* and the Barnes Foundation for providing a paint sample and for use of their image of *The Flower Piece*.

Author Contributions

BDAL acquired and analyzed TEM, STEM, XEDS, EELS, SAED and EMPAD data, and wrote MATLAB scripts for data processing. MGT used the FIB to produce thin sections of paint for TEM analysis. KXN assisted in acquiring EMPAD data and contributed MATLAB codes to process EMPAD data. MEH acquired EELS data, and aided in its interpretation. IG and UP were instrumental in the earliest phases of research on *The Scream* and have ownership of and collected the original paint samples. JLM and DAM initiated the project. BDAL, ACF, AMH, JLM and DAM and were involved in extensive discussions on the interpretation of the results, and in writing the manuscript.

Competing Interest Statement

The authors declare no competing interests.

FIGURES

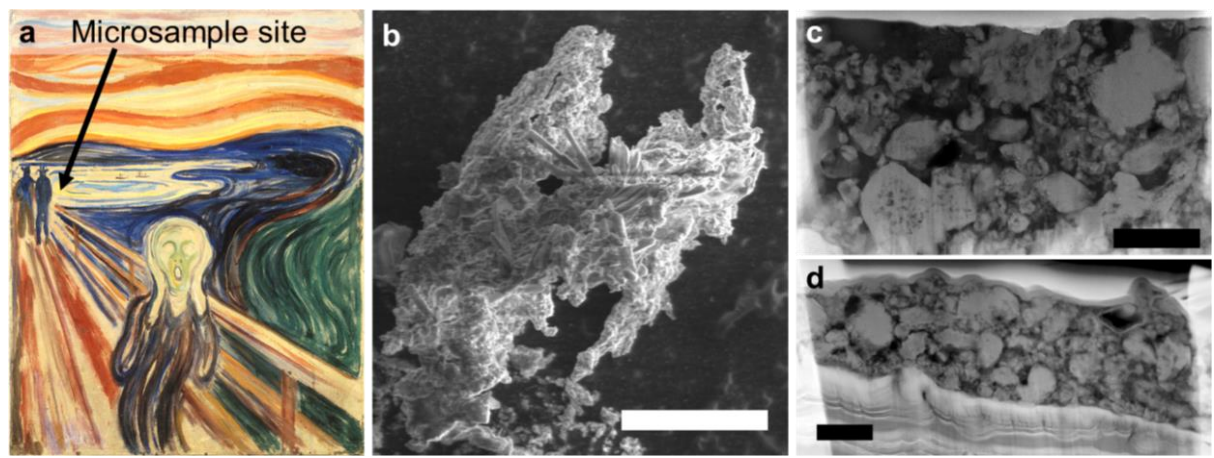

**Figure 1 | TEM sample preparation for paint from *The Scream*. a** Image of Edvard Munch's *The Scream* (c. 1910, Photo © Munch Museum). Four different versions of *The Scream* were created by Munch between 1893 and 1910. The 1910 version held by the Munch Museum, painted with mixed tempera/oil paints on cardboard is the final version. The arrow on the image shows the location from which the paint microsample analysed in this work was originally removed. **b** Scanning electron microscope (SEM) image of a typical paint flake in the sample prior to focused ion beam (FIB) milling. Scale bar 20 μm. **c-d** Transmission electron microscope (TEM) images of cross-sections of paint, referred to in the manuscript as "Scream 1" (c) and "Scream 2" (d), extracted by FIB milling. Scale bars 1 μm. Typically, FIB is used to thin TEM samples to <40 nm to ensure a high degree of electron transparency for high resolution imaging.[13] However, due to the fragility of the paint binding medium, we were limited to thinning our paint samples to ~100-150 nm to ensure that the paint cross section remained intact. The FIB milling procedure is described in greater detail in Extended Data Figure 1.

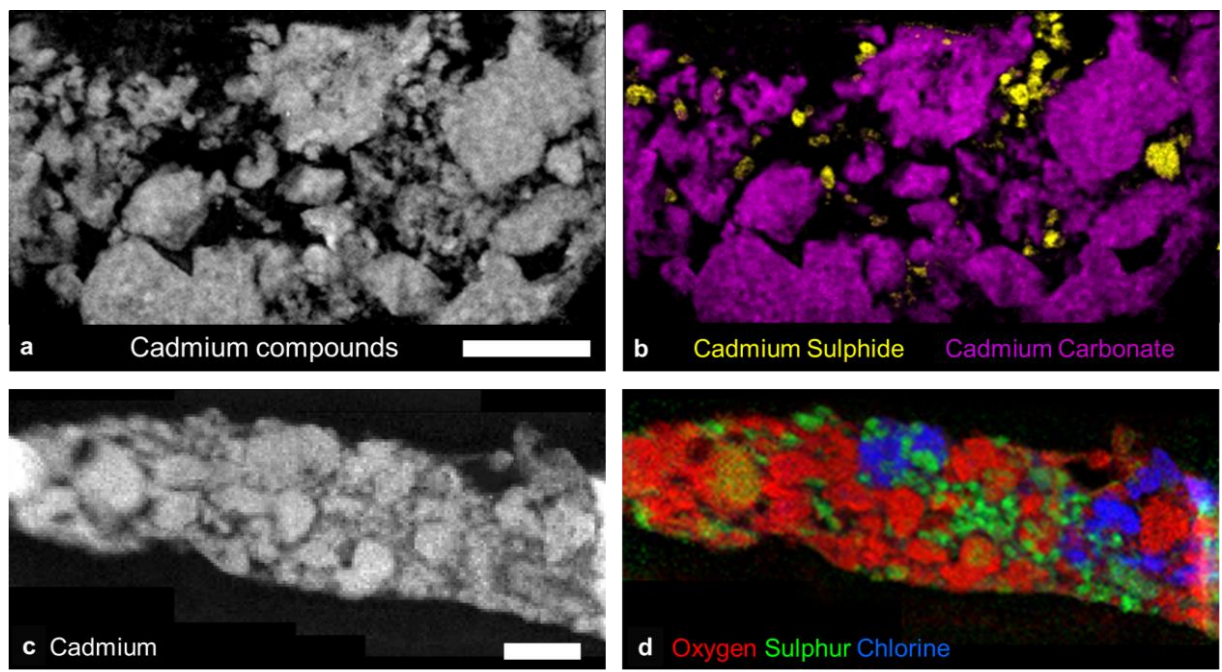

**Figure 2 | Spectroscopic analysis of paint samples from *The Scream* (c. 1910). a** Electron energy loss spectroscopy (EELS) map of the integrated cadmium M-edge in sample Scream 1, showing the distribution of all compounds containing cadmium. Scale bar 1 μm. **b** Colour overlay EELS map showing the distribution of different cadmium compounds in sample Scream 1. The cadmium sulphide (CdS) pigment, which comprises only a minority of paint particles, was identified where both the cadmium M-edge and sulphur L-edge are present, but other edges (such as the oxygen K-edge) are absent. Cadmium carbonate ($CdCO_3$) is identified by the fine structure of the carbon K-edge, and is mapped separately from other carbon compounds using a least squares fitting procedure, discussed in the supplementary information. **c** X-ray energy dispersive spectroscopy (XEDS) map of the integrated cadmium L-edge in sample Scream 2. Scale bar 1 μm. **d** Colour overlay showing the distribution of oxygen (red), sulphur (green) and chlorine (blue) in sample Scream 2. Here, the sulphur signal is associated with CdS. As in sample Scream 1, CdS only comprises a minority of paint particles. The chlorine signal (blue) is mainly associated with cadmium chloride. Additionally, a small chlorine peak is also seen X-ray spectra from CdS particles, implying chlorine doping of CdS

(Extended Data Figure 2). Additional EELS measurements suggest the oxygen signal is largely associated with $CdCO_3$ (Extended Data Figures 4 and 5).

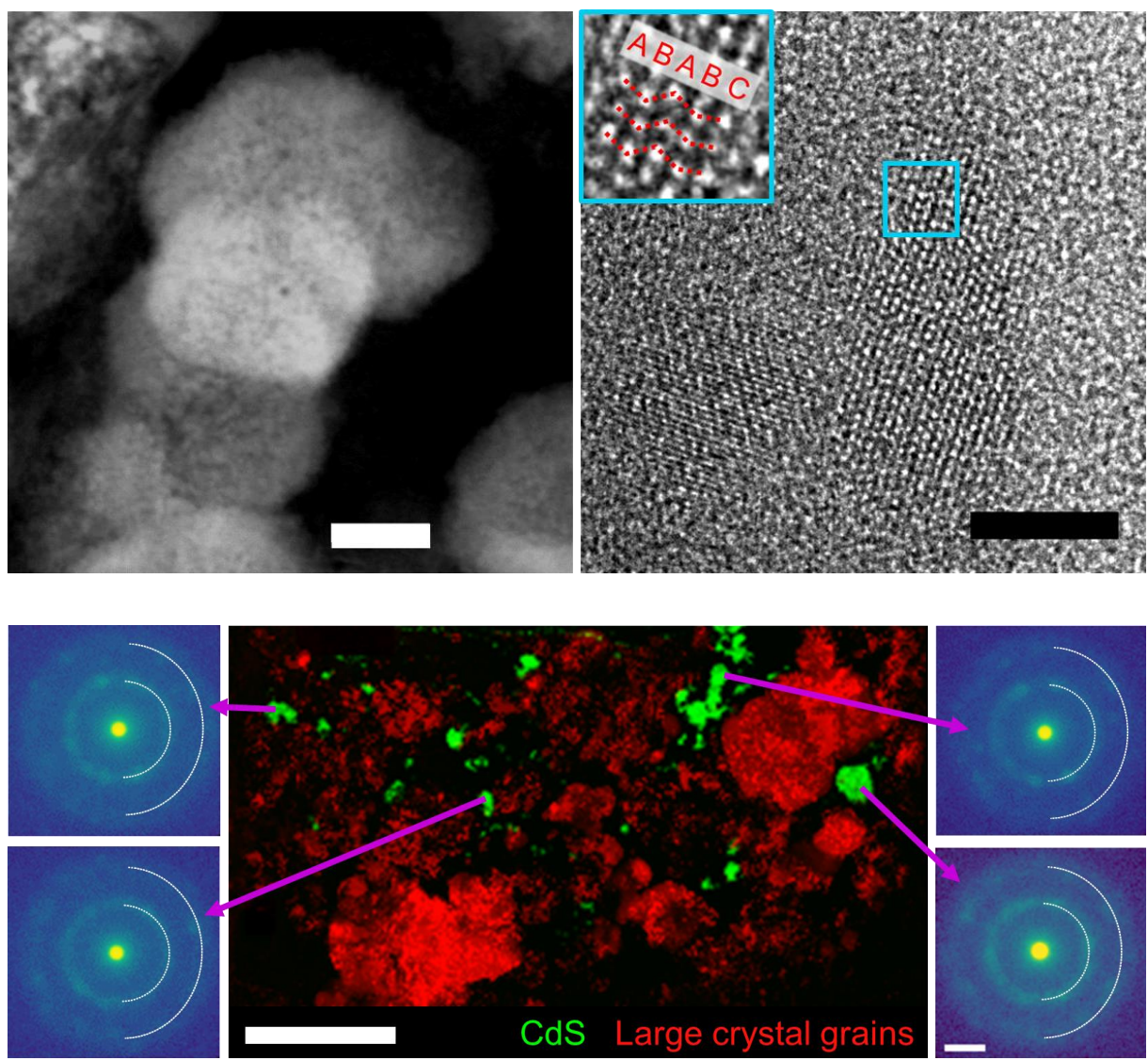

**Figure 3 | High resolution imaging and diffraction mapping of pigment particles. a** High angle annular dark field (HAADF) scanning transmission electron microscope (STEM) image of an area of cadmium sulphide (CdS) pigment in sample Scream 1. The image suggests that the CdS consists of clusters of nanoparticles of ~ 5-10 nm in diameter. Scale bar 50 nm. **b** Lattice resolution zero-loss energy-filtered TEM (EFTEM) image of CdS nanoparticles of 5-10 nm diameter in sample Scream 1. Scale bar 5 nm. The inset in d highlights an "ABABC" stacking sequence in the particle in the region indicated by the light blue box. These images have been Wiener filtered to improve signal to noise. **c** The central panel shows a red/green colour overlay of the map of large (>10 nm) crystal grains in the sample with the CdS EELS

map from Figure 2b. Areas of significant overlap, indicating the presence of large crystal grains would appear yellow in the figure, but no yellow regions are present. Individual convergent beam electron diffraction (CBED) patterns from CdS regions are shown in the side panels and all show a polycrystalline ring pattern. White dashed lines are added to aid the reader in identifying the rings. This is consistent with the nanoparticle structure that we observe in our images. The scale bar on the central panel is 1 μm, and the scale bar shown on the CBED pattern is 2.5 nm$^{-1}$. Further HAADF-STEM and EFTEM images are given in Extended Data Figures 6 and 7, and scanning diffraction data for sample Scream 2 is shown in Extended Data Figure 8.

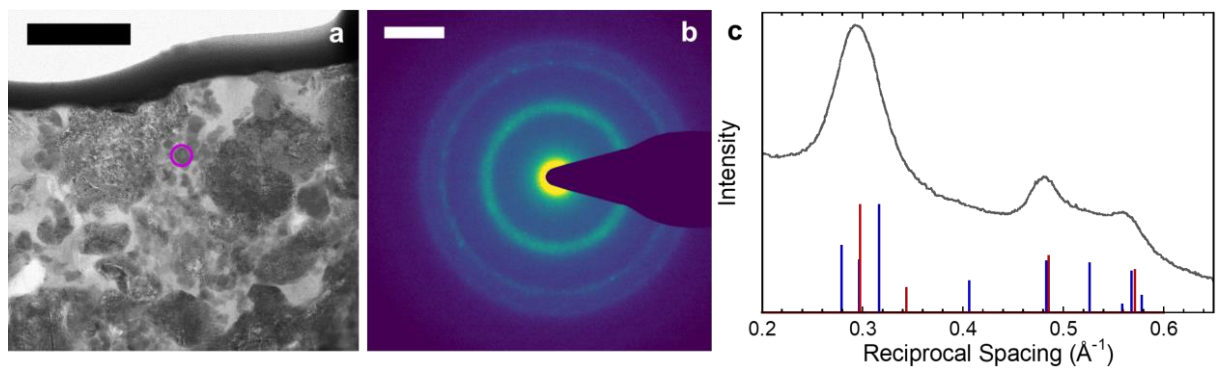

**Figure 4 | Analysis of selected area electron diffraction patterns from cadmium yellow pigment particles. a** Transmission electron microscope (TEM) image of part of sample *Scream 1*. Scale bar 1 µm. **b** Selected area electron diffraction (SAED) pattern from a region of CdS indicated by the magenta circle in the TEM image (a). Scale bar 0.25 nm$^{-1}$. **c** Rotational average of the SAED pattern (b). The positions and relative intensities of standard greenockite (blue) and hawleyite (red) peaks are indicated by the lines under the diffraction pattern. These diffraction patterns are not a precise match to either greenockite, or hawleyite, which suggests that the CdS nanoparticles may have a polytype crystal structure.[21]

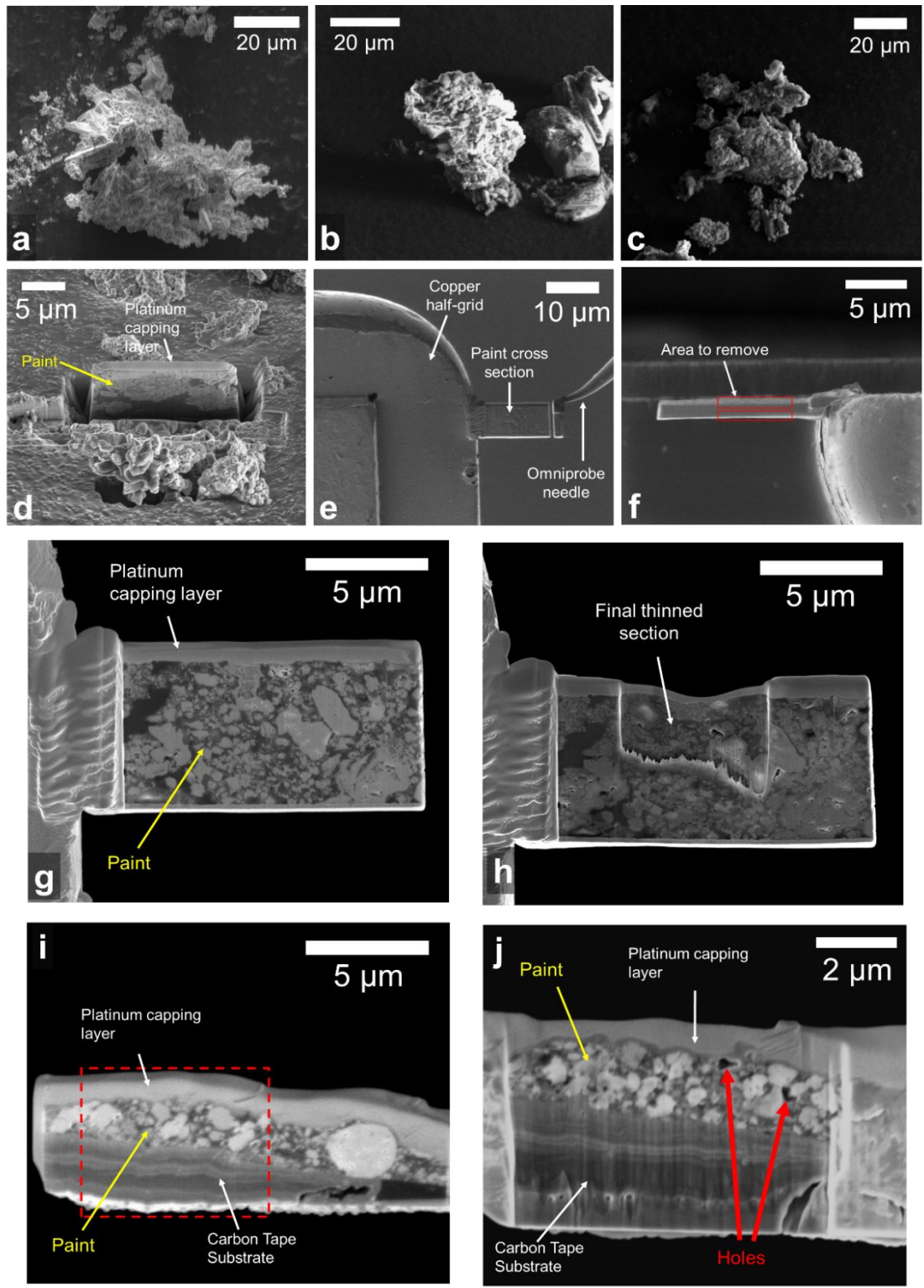

**Extended Data Figure 1 | FIB process for extracting a thin slice of pain from a paint flake.**.

**a-c** SEM images of paint flakes in the sample of paint from *The Scream*. **d** Example image of a section of paint capped by a protective platinum layer, with the area surrounding the section milled away. **e-f** Paint cross section extracted with Omniprobe needle and welded to copper half-grid with platinum. **g-h** Sample Scream 1 before (g) and after (h) FIB thinning. **i-j** Sample Scream 2 before (i) and after (j) FIB thinning.

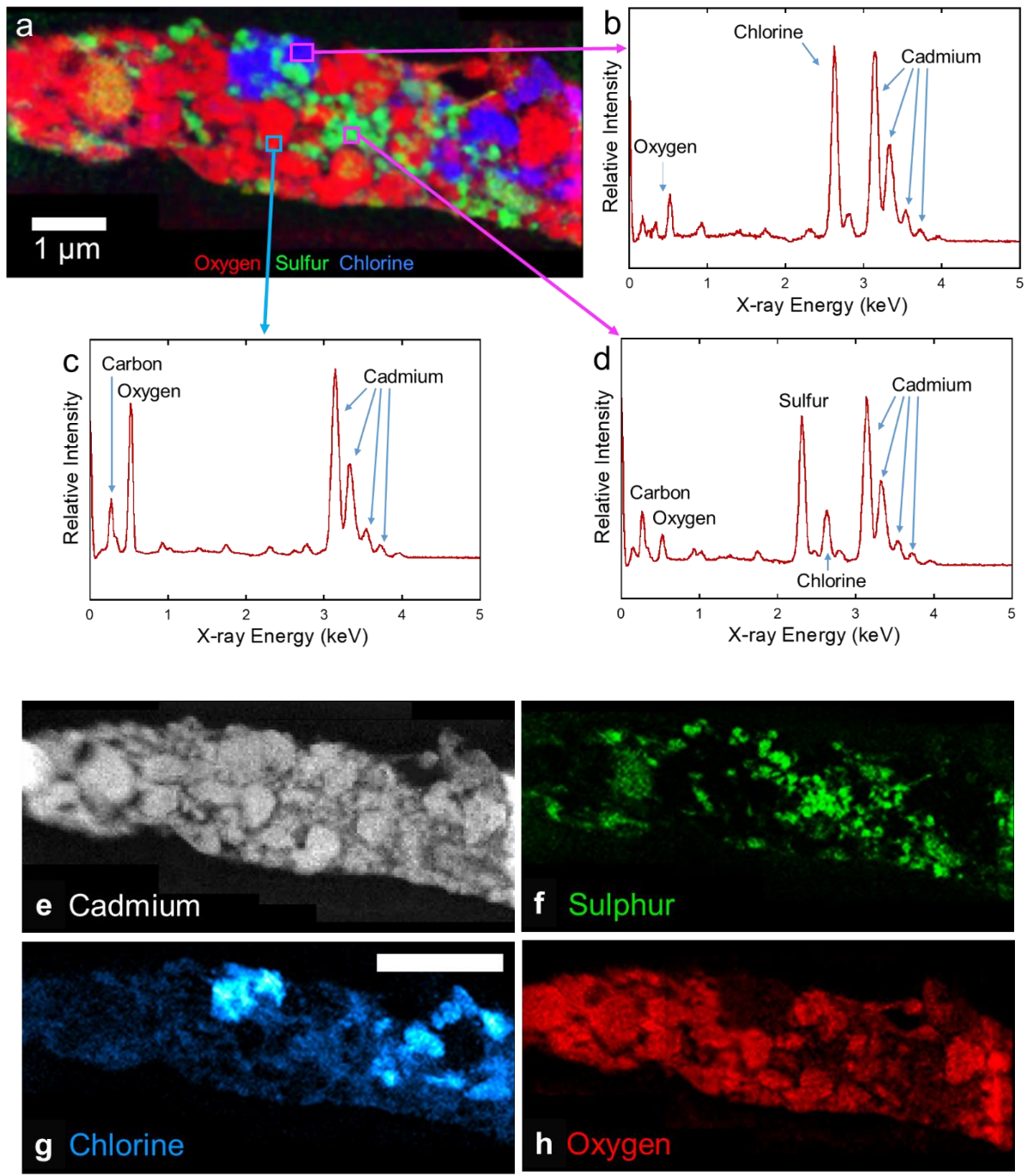

**Extended Data Figure 2 | XEDS spectra and maps from different regions of sample Scream 2.** X-ray spectra contain quantitative information about the relative concentration of elements in different parts of a sample. This information is not always apparent from XEDS maps alone. **a** A colour overlay map showing the relative concentrations of oxygen, sulphur

and chlorine, for reference. **b** X-ray spectrum from chlorine and cadmium rich particles. A weaker concentration of oxygen is also observed in this region. These chlorine rich regions may be residual starting reagent from CdS synthesis. **c** X-ray spectrum from cadmium and oxygen rich particles. Carbon, oxygen, and cadmium are the only major X-ray signals here. **d** X-ray spectrum from sulphur and cadmium rich particles. Weaker concentrations of carbon, oxygen, and chlorine are also observed in this region. The weak concentration of chlorine may indicate chlorine doping of CdS particles. **e-h** Elemental maps of the signals from cadmium, sulphur, chlorine, and oxygen.

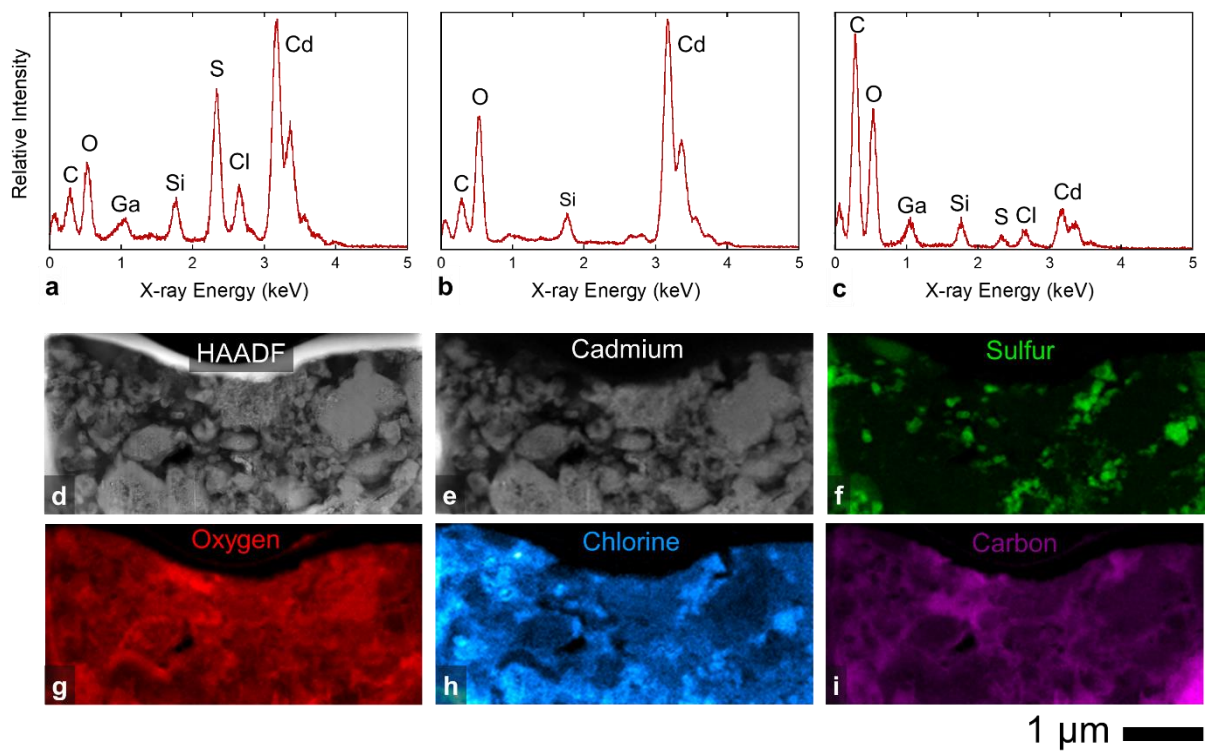

**Extended Data Figure 3 | XEDS maps and spectra from sample Scream 1.** X-ray spectra from representative sets of particles have been chosen to show more information about the elemental composition of the particles in the paint. **a** Typical XEDS spectrum from a paint particle containing high quantity of sulphur. As noted in Extended Data Figure 2d, the presence of chlorine in this region may be consistent with chlorine doping of cadmium sulphide particles. **b** Typical XEDS spectrum from a paint particle containing high quantity of oxygen. **c** X-ray spectrum from a region of binding medium, containing no paint particles. This area has a high concentration of carbon and oxygen. Weak concentrations of sulphur, chlorine, and cadmium, may suggest some leaching of these elements into the binder. **d** HAADF image of sample Scream 1 tilted to 30° for XEDS acquisition. **e-i** XEDS elemental maps from sample Scream 1. Comparing each elemental map to the HAADF image (d), it is clear that most paint particles contain cadmium (e), but only a minority contain sulphur (f). Oxygen (g), chlorine (h), and carbon (i) are present in varying concentrations in the paint particles and the binder. It should

be noted that in addition to X-ray signals arising from the paint, the gallium L-edge (~1.10 keV) is present in our X-ray spectra due to gallium deposition from FIB preparation. A silicon K-edge signal (~1.74 keV) is present due to use of a Si(Li) X-ray detector, and the copper L-edge (~0.93 keV) is present due to scattering of some electrons into the copper half-grid holding the sample.

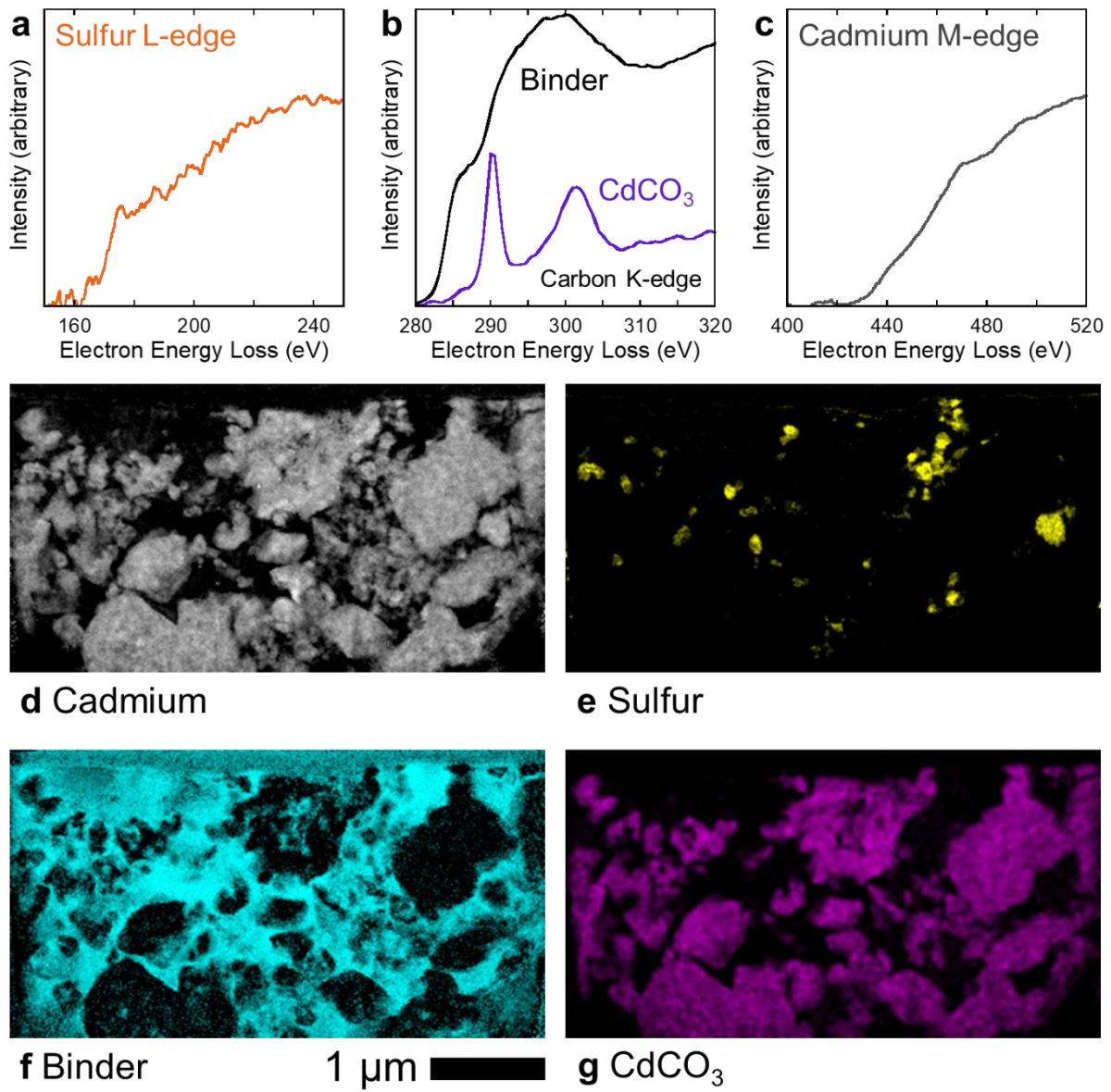

**Extended Data Figure 4 | EELS spectra and maps from sample Scream 1** Background subtracted EELS edges **a** Sulphur L-edge. **b** Carbon K-edge. **c** Cadmium M-edge. Each of these spectra has been averaged over a total area of the sample of at least 100 nm$^2$. **d-g** EELS maps of FIB cross section "Scream 1", showing the distribution of cadmium (d), sulphur (e), binder (f), and cadmium carbonate (g). These EELS maps are consistent with thee data from our XEDS maps (Extended Data Figure 3), but provide additional information regarding carbon bonding. Separate maps of cadmium carbonate and the binding medium were generated using

least squares fitting (see Methods). Comparing the cadmium map to the $CdCO_3$ map, one can see that a majority of paint particles are $CdCO_3$. This observation is consistent with previous bulk measurements that have suggested that cadmium yellow paint in *The Scream* contains a high concentration of $CdCO_3$[2].

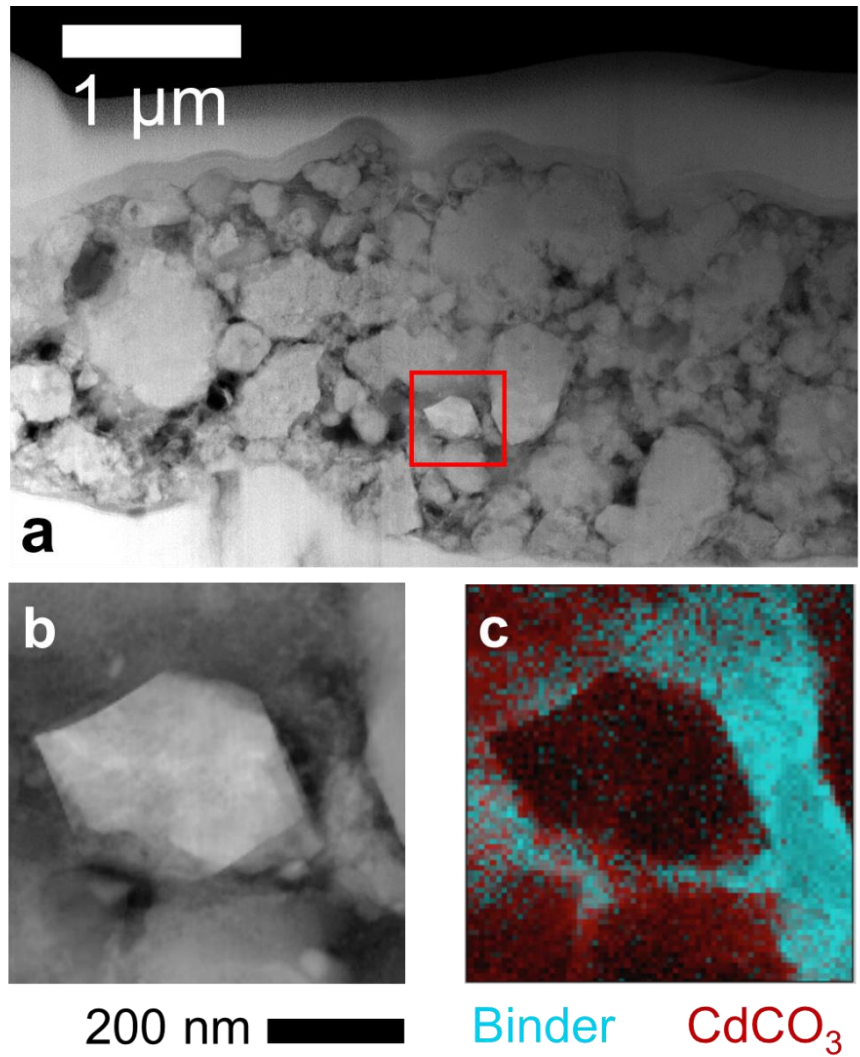

**Extended Data Figure 5 | EELS maps from sample Scream 2 a** HAADF image of part of sample Scream 2, the thicker of our two samples. Only select regions of Scream 2 proved thin enough for EELS mapping, even when using a 300 kV electron beam voltage. **b** Magnified HAADF image of the area marked by the red box in (a), which we were able to examine with EELS. The area had been established as cadmium and oxygen rich in XEDS. In the area surrounding the particle, we were able to distinguish binder and $CdCO_3$ in the same way as in sample Scream 1 (Extended Data Figure 4b). This indicated that $CdCO_3$ was also present in sample Scream 2. **c** Binder and $CdCO_3$ overlay EELS map of the area in (b).

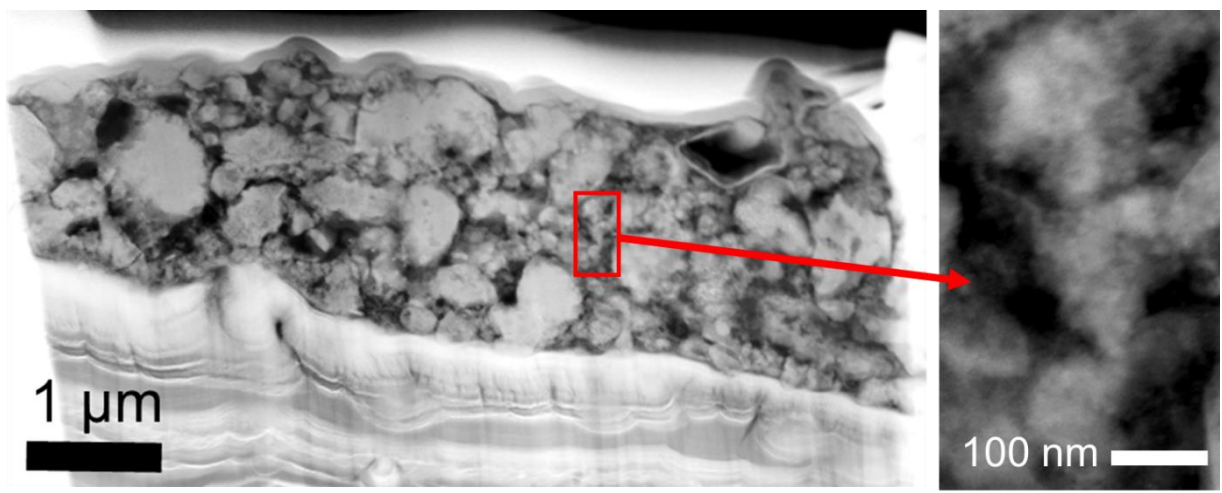

**Extended Data Figure 6 | HAADF images CdS pigment particles in sample Scream 2.** HAADF-STEM images of regions of CdS in sample Scream 2, reveal a granular structure, similar to that observed in sample Scream 1 (Figure 3). This suggests that the CdS in the paint is composed of clusters of small (< ~10 nm) nanoparticles.

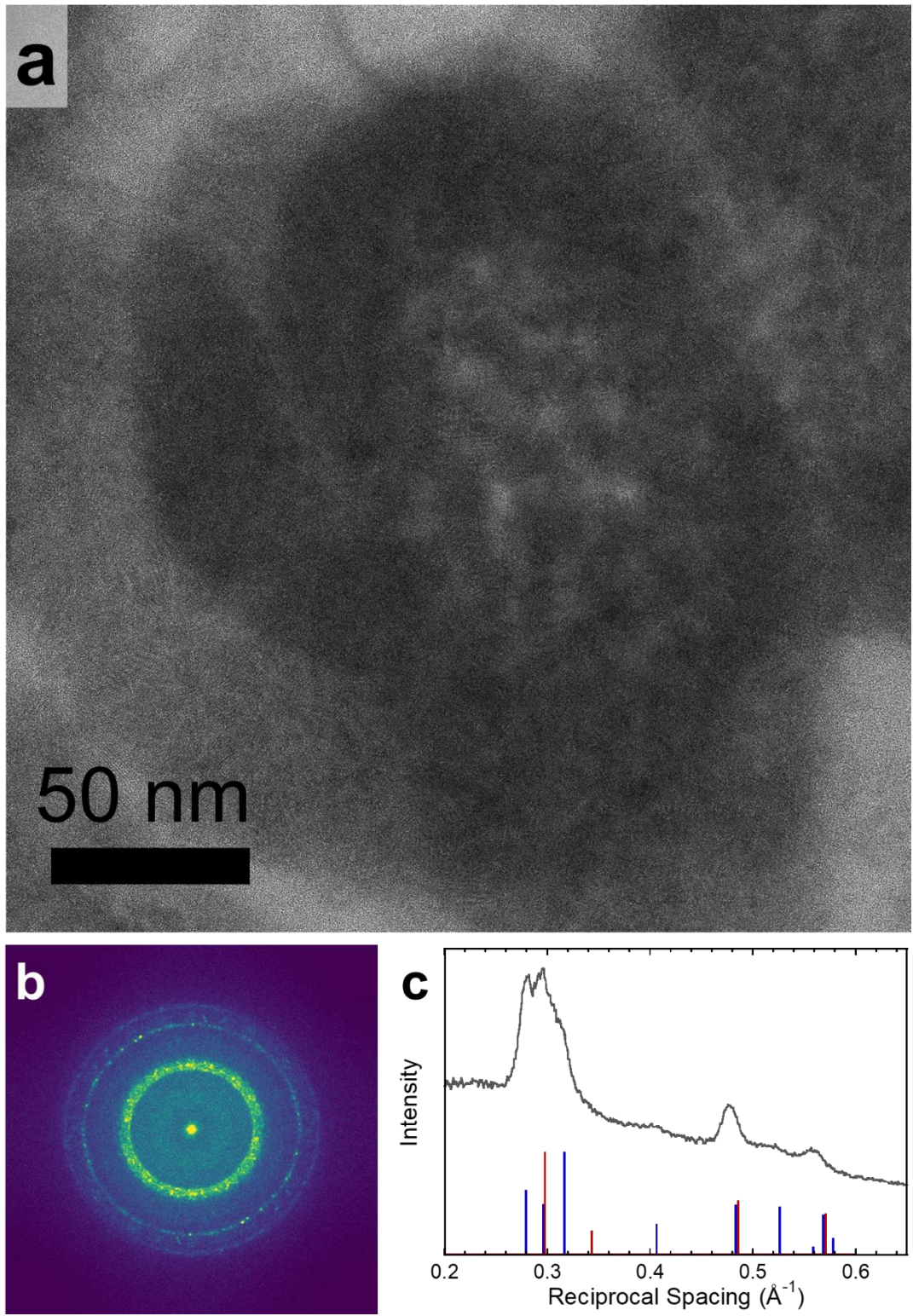

**Extended Data Figure 7 | Fourier transform analysis of EFTEM images** Fourier transforms of large field of view images with lattice resolution provide complimentary to information to SAED. **a** Image of a cluster of CdS particles **b** Fast Fourier transform of the image in (a). **c** A rotational average of the Fourier transform in (b). The positions of peaks associated with

hexagonal and cubic CdS are shown below in blue and red. The FFT is very similar to our SAED data, which is consistent with a polytype crystal structure (Figure 4).

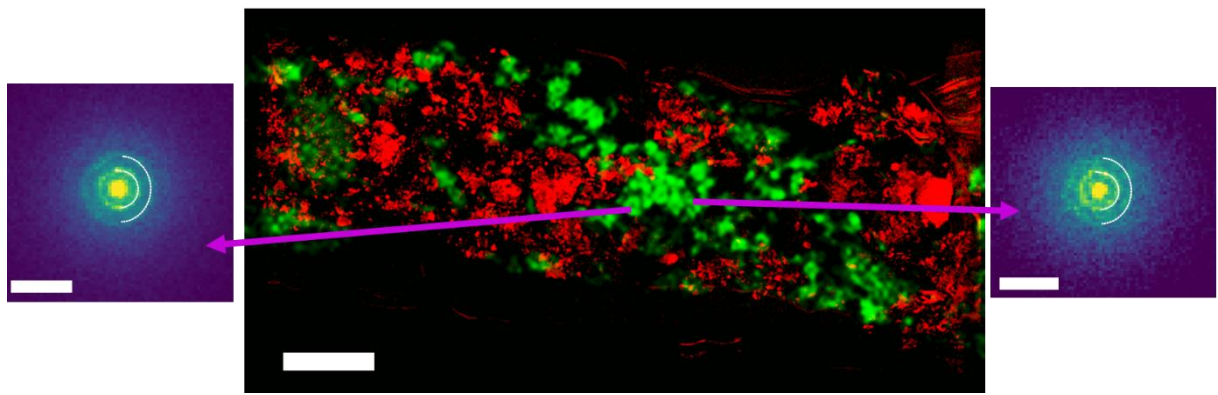

**Extended Data Figure 8 | Crystallographic Mapping of samples Scream 1 and Scream 2.**

The central panel shows a red/green colour overlay of the map of large (>10 nm) crystal grains in sample Scream 2 with the sulphur component of the XEDS map from Extended Data Figure 2. This map was derived from a 4D CBED dataset in the same manner as in Figure 3c. Areas of significant overlap, indicating the presence of large CdS crystal grains, would appear yellow in the figure, but no yellow regions are present. CBED patterns from CdS regions are shown in the side panels and show a polycrystalline ring pattern. This suggests that the CdS is nanocrystalline, as was the case in Figure 3c. The scale bar on the central panel is 1 μm, and the scale bar shown on the CBED pattern is 10 nm$^{-1}$. The diffraction rings are consistent with those observed in the CBED patterns in Figure 3c, but the diffraction patterns were acquired using a different camera length, meaning that the scale bar is different.

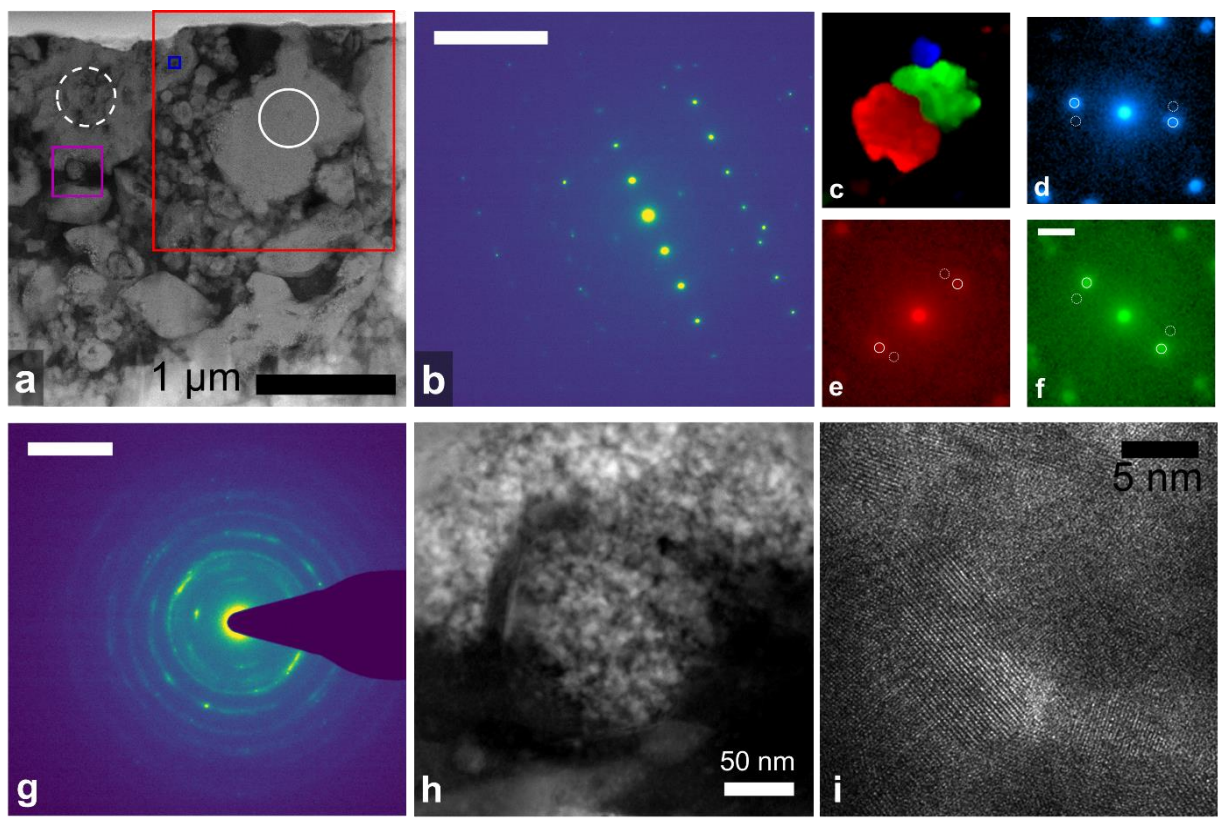

**Extended Data Figure 9 | Variation in CdCO$_3$ crystal grain size in *The Scream*. a** ADF-STEM image of part of the sample. Scale bar 1 µm **b** A SAED pattern from the area indicated by the solid white circle in (a). **c** A red/green/blue colour overlay map showing the three large crystal grains in the area indicated by the red box in a. **d-f** Colour coded CBED patterns for each of the crystal grains shown in (b). Scale bar 2.5 nm$^{-1}$. For each CBED pattern, an image of the associated grain is produced from the EMPAD data by generating centred dark-field images from virtual apertures placed over diffraction spots (solid white circles). In this case, these spots all match the (104) d-spacing of CdCO$_3$. Background noise is reduced by subtracting images from virtual apertures placed off the diffraction spots (dashed white circles). Given the large size (>200 nm) of the three grains shown in **c**, we consider it unlikely that this particle formed due to the aging of <10 nm CdS nanoparticles. We hypothesize that CdCO$_3$ particles with large crystal grain sizes were originally present in the paint either as unreacted starting reagent from CdS synthesis, or due to the addition of CdCO$_3$ as an extender. **g** SAED

pattern from the area indicated by the dashed white circle in (a), with a ring pattern indicating many crystal grains in the area This region was found to consist mainly of $CdCO_3$ in spectroscopic mapping (Figure 2). Scale bar 5 nm$^{-1}$. **h** ADF-STEM image from area of $CdCO_3$ cluster indicated by magenta box in (a), scale bar 50 nm, and **i** Zero-loss EFTEM image of area of $CdCO_3$ cluster indicated by blue circle in (a). Both show nanoparticles of a similar size to the CdS nanoparticles imaged in Figure 3.

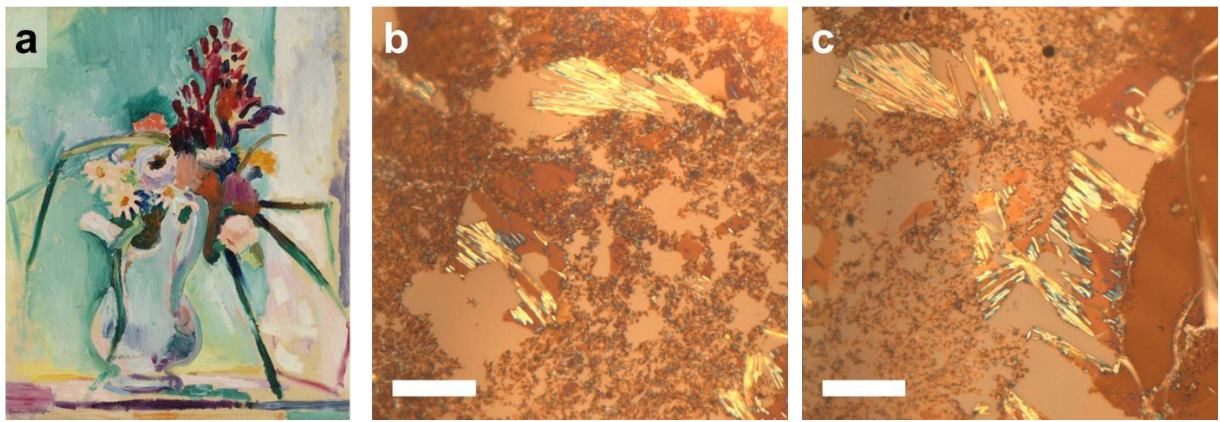

**Extended Data Figure 10 | Microtomed sections of cadmium yellow paint from Henri Matisse's *The Flower Piece*. a** Henri Matisse's *Flower Piece* (1906, The Barnes Foundation). **b** and **c** Optical microscope images of areas of a microtomed paint cross section. Scale bars 20 µm. The bright yellow regions are areas of cadmium yellow paint. Many tears and holes are visible in the optical images microtomed cross section, which suggests that some material has fallen out of the cross section. However, the bright yellow regions suggest that CdS pigment particles are still present in the sample.

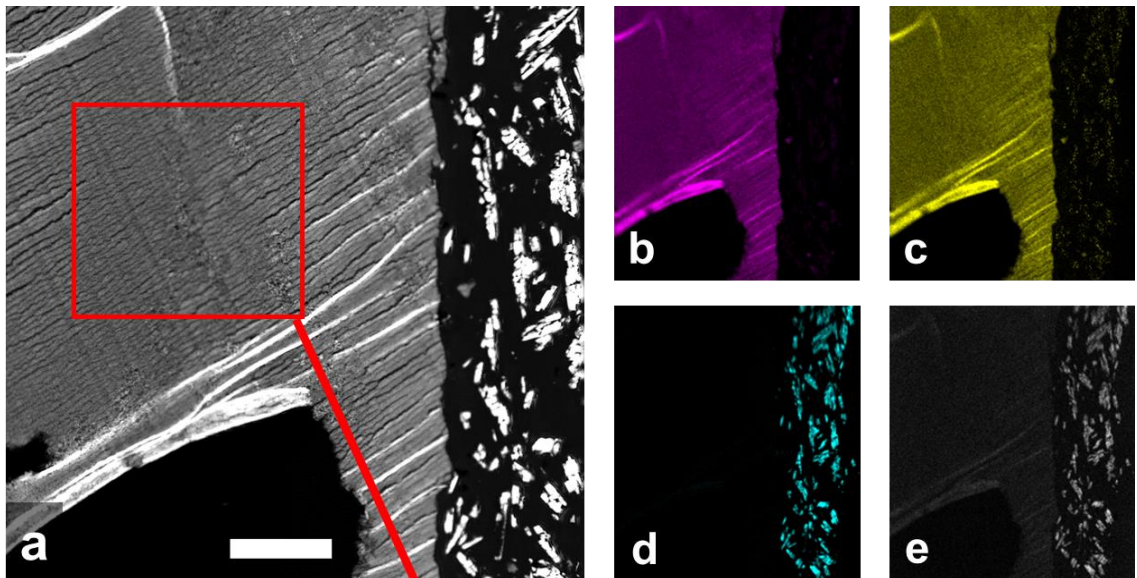
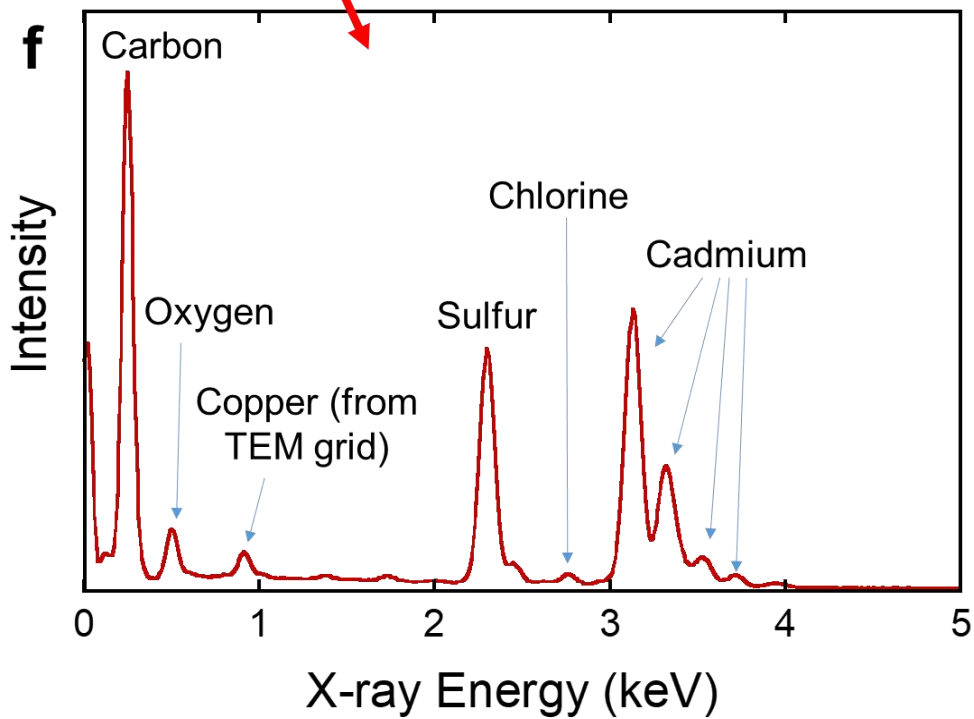

**Extended Data Figure 11 | Analysis of microtomed cadmium yellow paint from Henri Matisse's *The Flower Piece*. a** ADF-STEM image of part of a microtomed section of paint. Scale bar 2 µm. XEDS maps of this area show the distribution of **b** cadmium, **c** sulphur, **d** lead, and **e** oxygen. The small particles to the right containing lead are likely to be lead white paint particles ($PbCO_3$). **f** The sum spectrum of an area of paint shown by the red box in (a). The strong carbon signal is due to the embedding epoxy. The strong cadmium and sulphur signals

from this region suggest a predominantly cadmium sulphide composition. A small chlorine peak is observed, suggesting that the CdS particles may be chlorine doped. Overall, the XEDS, data suggests a similar elemental composition to the CdS pigment in *The Flower Piece* as that observed in *The Scream,* although we note that the ratio of the chlorine peak relative to the sulphur peak appears to be smaller for the *Flower Piece*.

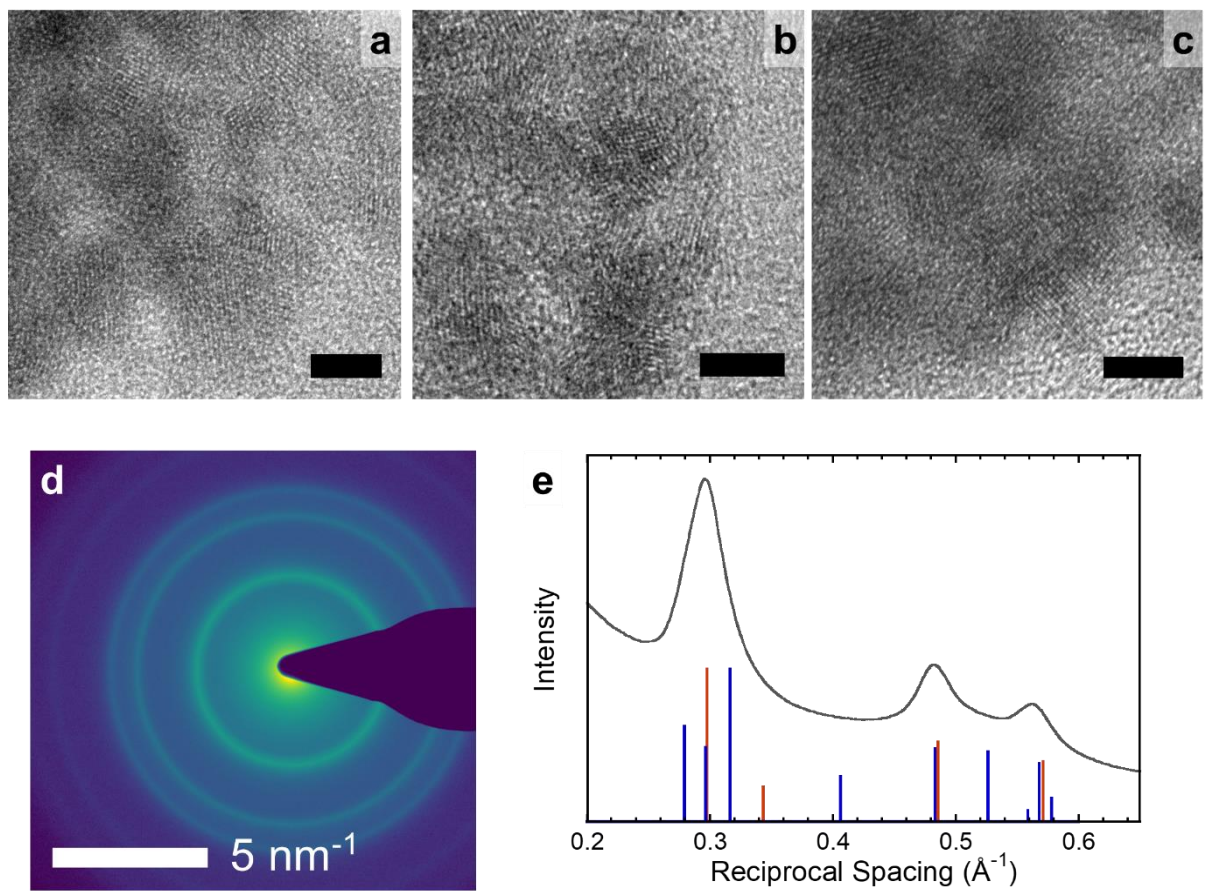

**Extended Data Figure 12 | Analysis of microtomed cadmium yellow paint from Henri Matisse's *The Flower Piece*. a-c** HRTEM images of CdS particles in microtomed section of paint from The Flower Piece. As observed in Extended Data Figure 10, the microtomed section of paint contained a number of stretches and tears that occurred as the section was cut. Near to the edges of tears and holes in the microtomed section, CdS pigment particles can be imaged at lattice resolution in TEM without using an energy filter. Typical CdS particle size is in the 5 – 10 nm range, similar to that observed in *The Scream*. Scale bars 5 nm. **d** A selected area diffraction pattern from an area of CdS particles. The pattern shows polycrystalline rings. Scale bar 5 nm$^{-1}$. **e** Rotationally averaged SAED pattern from CdS particles in *The Flower Piece*. The positions and intensities of greenockite and hawleyite peaks are shown by the blue and red lines respectively. The presence of the three broad diffraction peaks, suggest that the CdS is

composed of polytype nanoparticles. Again, this is very similar to that observed for CdS in *The Scream*.

**Supplementary Discussion - CdCO$_3$ particle sizes and relationship to photodegradation**

It has been proposed that CdCO$_3$ may be the end product of CdS photodegradation.[1] Indeed, the conversion of yellow CdS into white CdCO$_3$ is consistent with the fact that the paint has been observed to whiten with age[1-3]. Much of the paint in the samples from *The Scream* consists of CdCO$_3$ (Figure 2b). However, the detection of CdCO$_3$ alone does not constitute proof of CdS photodegradation, because CdCO$_3$ may also have been originally present in the paint as a residual starting reagent from CdS synthesis, although this is a more unlikely scenario because of the cost of the material.[1] We can collect information about the size and of the CdCO$_3$ particles to help resolve this ambiguity. Using both electron diffraction and high resolution EFTEM imaging, we have observed that some CdCO$_3$ particles in the paint are ~1 μm in size, whilst others are ~10 nm in size (Extended Data Figure 9), which is similar to the size of the CdS particles. Given this difference in size, we hypothesize that there may be two sources of CdCO$_3$ in *The Scream*'s yellow paint. The larger, more well-faceted CdCO$_3$ particles, with a relatively low surface area to volume ratio, and therefore lower reactivity, may have originally been present in the paint. This hypothesis is supported by ongoing studies of Edvard Munch's cadmium yellow paint tubes, which have in fact yielded CdCO$_3$ even in tubes that have remained sealed from the environment. However, we consider that small CdCO$_3$ nanoparticles that are similar in size to the observed CdS nanoparticles in the pigment are likely to be CdS aging products. Conversion of yellow CdS nanoparticles to white CdCO$_3$ nanoparticles would be consistent with the whitening of the paint, which has been observed to occur over time.

References for Supplementary Discussion

1. Mass, J.L., Opila, R., Buckley, B., Cotte, M., Church, J. & Mehta, A. The photodegradation of cadmium yellow paints in Henri Matisse's Le Bonheur de vivre (1905-1906), *Appl. Phys. A,* **111** (1), 59–68 (2013a).